\documentstyle[12pt]{article}

\input epsf

\begin{document}
\begin{titlepage}
\today          \hfill
\begin{center}
\hfill    OITS-673 \\
\hfill    LBNL-43082 \\
\hfill    UCB-PTH-99/12
\vskip .05in

{\large \bf Signals of Supersymmetric Lepton Flavor Violation at the
LHC} \footnote{A preliminary version of this work 
was presented by KA at the Higgs and 
Supersymmetry: Search and Discovery Conference at the University of Florida,
Gainesville, March 8-11, 1999.}$^,$
\footnote{This work is supported by DOE Grant DE-FG03-96ER40969.}
\vskip .15in
Kaustubh Agashe $^{a,}$\footnote{email: agashe@oregon.uoregon.edu.} 
and Michael Graesser  $^{b,}$\footnote{email: mlgraesser@lbl.gov.}$^,$ 
\footnote{Supported by the Natural Sciences and Engineering Research 
Council of Canada.}
\vskip .1in
$^a$ {\em
Institute of Theoretical Science \\ 
 5203 University
of Oregon 
Eugene OR 97403-5203} \\
\vskip .1in
$^b$ {\em
Theoretical Physics Group \\
Lawrence Berkeley National Laboratory \\
University of California \\
Berkeley CA 94720}
\end{center}

\vskip .05in

\begin{abstract}
In a generic supersymmetric extension of the Standard Model, there will
be lepton flavor violation at a neutral gaugino vertex
due to 
misalignment between the lepton
Yukawa couplings and the slepton soft masses.
Sleptons produced at the LHC through the 
cascade decays of squarks and gluinos can
give a sizable number of events with $4$ leptons.
This channel could
give a clean signature of supersymmetric lepton flavor violation 
under conditions which are identified.
\end{abstract}

\end{titlepage}

\renewcommand{\thepage}{\arabic{page}}
\setcounter{page}{1}

\section{Introduction}

In the supersymmetric Standard Model (SM), the quadratically divergent
corrections to the Higgs (mass)$^2$ cancel due to supersymmetry (SUSY).
The remaining corrections are logarithmically divergent, proportional
to the SUSY breaking masses of the sparticles (the superpartners of the 
SM particles) and result in
a negative Higgs (mass)$^2$ due to the large top quark Yukawa coupling.
Thus, the superpartners of the SM particles 
must have masses $\stackrel{<}{\sim} 1$ TeV
in order for
SUSY to solve the gauge hierarchy problem and lead to natural electroweak
symmetry breaking.

With 
the sparticle masses at the weak scale, 
these new particles
(especially gluinos and squarks) 
will be produced in significant amounts at the
LHC.
After the initial discovery of the sparticles,
the focus will be on precision measurements 
of their masses and mixings just as, for
example, the next step after the discovery of the heavy quarks
was the measurement of their detailed properties. 
In this paper, a relatively clean signal at the LHC for 
detecting 
the mixing angle between the
scalar partners of the charged leptons (the sleptons) is presented.

A flavor-violating signal is obtained from the production of 
real sleptons, followed by their oscillation into a different 
flavored slepton, and subsequent decay to a lepton.
Some formulae for these oscillations are given in section \ref{formulae}.
At a $e e$ linear collider, the production of slepton pairs
can then give $e \mu$ events with missing energy. This 
was studied in \cite{krasnikov0,nima1}. Dilepton flavor {\em and} CP violating 
signals at the LHC and NLC were studied in \cite{nima2}. 
At a hadron collider (the LHC), sleptons can be pair-produced by the 
Drell-Yan process
giving the same signal. This was studied in \cite{krasnikov2,krasnikov1}, 
and is a promising signal for large flavor mixing angles and when the 
SUSY background is known to be small. 
Real sleptons can also be produced at the LHC
in the 
decays of the next-to-lightest neutralino $(\chi^0 _2)$, which 
are mainly produced in the
cascade decays of gluinos and squarks.  
In section \ref{cascade2l}, flavor violating
dilepton events from $\chi_2^0$ decays are briefly
considered.
The production of $\chi^0_2$ {\em pairs} can 
give rise to events with 4 leptons, with  
the dramatic flavor violating 
signal identified by a $(3e+\mu)$ or $(3\mu +e)$ lepton signature, 
hard jets, no $b-$jets, and of course 
missing energy. This is discussed in section \ref{cascade4l}.  
Conditions on the supersymmetric 
spectrum that are favorable for the suppression of the 
dominant supersymmetric background, 
occuring from 
{\em heavier} neutralino/chargino and stop decays,
are identified. 
Ideas for determining the remaining  
dominant supersymmetric 
background occuring from $\tau$ decays are also presented. 
These are all conveniently  
summarized in the Conclusion. 
 
In section \ref{quick}, a brief estimate of the expected 
$4$-lepton signal at
a generic point in SUSY parameter space is given.
Next, in section \ref{LHCpoint},
a particular LHC Point \cite{snowmass,ian} 
is considered.  It is found that at this Point, a $5\sigma$ discovery 
($2\sigma$ exclusion) is obtained for a right-handed (RH) 
first and second generation 
mixing angle $\sin  \theta_R >0.13$ $(\sin \theta_R >0.08)$ 
with an integrated luminosity of $100$ fb$^{-1}$ at {\em low luminosity}.
The discovery potential at high luminosity is still optimistic 
though less quantitative, due to uncertainties
in $\tau-$jet detection efficiencies and larger $b-$jet mistagging rates.  
In any case, 
the values for the mass splitting 
(between $\tilde{e}$ and $\tilde{\mu}$)
that are favorable for the discovery 
of a signal satisfy the $\mu \rightarrow e \gamma$ bound even 
for a maximal 
mixing angle. Thus the LHC has the opportunity of probing mixing angles 
that are beyond the reach of the current $\mu \rightarrow e \gamma$ 
limit.

\subsection{Lepton Flavor Violation due to Slepton Mass Mixing}
\label{formulae}
To begin consider the lepton-slepton-neutral gaugino vertex with the leptons
and sleptons in the gauge basis:
\begin{equation}
\tilde{l}_{a}^{\ast \; gauge} \; l^{gauge}_{a}
\chi^0,
\label{basis1}
\end{equation} 
where $a =1,2,3$ is a flavor index.
Next perform a
unitary transformation, $V$, on both $l_g$ {\em and} $\tilde{l}_g$ to 
go to the mass eigenstate
basis for the $l$'s:
\begin{equation}
\tilde{l}_{\alpha}^{\ast} l_{\alpha} \chi^0.
\label{basis2}
\end{equation}
In this basis the coupling remains diagonal in 
flavor space (now denoted
by $\alpha$).
In general, 
however, the slepton and lepton mass
matrices are {\em not} related so that
the same unitary matrix, $V$, may not diagonalize them {\em both}. 
In this general case,  
the slepton (mass)$^2$ matrix in the basis $\tilde{l}_{\alpha}$ is
{\em not} diagonal. For example, even if the slepton (mass)$^2$
matrix in the gauge basis $\tilde{l}_a^{gauge}$ of Eqn.(\ref{basis1})
is diagonal but not
$\propto$ {\bf 1}, it will have off-diagonal elements in the basis
$\tilde{l}_{\alpha}$ of Eqn.(\ref{basis2}).
So,
a further unitary transformation, $W$,  is needed 
to rotate to the slepton mass basis. In this basis the 
slepton-lepton-gaugino vertex is: 
\begin{equation}
\tilde{l}_{i}^{\ast} \; l_{\alpha}  W_{i \alpha} \chi^0. 
\end{equation}
So, in the mass basis for leptons {\em and} sleptons
($l_{\alpha}$ and $\tilde{l}_i$) 
a mixing matrix $W \neq$ {\bf 1} in general 
appears at the
neutral gaugino-lepton-slepton vertex.
This means that there is a coupling between, for example,
$\tilde{e}$ (in the mass basis), 
$\mu$ and $\chi ^0$ -- this will be referred to as SUSY
lepton flavor violation. 
The focus of this paper is the detection of 
this SUSY lepton flavor violation at the LHC.

The theoretical expectations for $W$ are varied. In  
models with broken flavor symmetries, it is expected that 
$W \sim V_{KM}$. In such cases a Cabibbo-like mixing angle for 
the first two generations and a $\Delta m/m $ close to the 
$\mu \rightarrow e \gamma$ bound is expected \cite{flavor}. 
In contrast, in models of 
gauge-mediated supersymmetry breaking  
the dominant contribution 
to the soft masses is universal and it naively appears that 
there is no interesting flavor physics. There is, however, a subdominant 
flavor non-universal supergravity contribution. This likely results in 
{\em large} mixing angles \cite{nima2}. The magnitude of $\Delta m/m$ depends 
on the supersymmetry breaking scale and while clearly model--dependent, 
could easily be $\sim \Gamma /m$ or larger, 
which is needed to give an observable flavor-violating signal at the LHC
(this is discussed later in this section). 

For simplicity, the 
case of $1-2$ mixing with mixing angle $\theta$ is discussed.
In this case there are strong limits on the mixing angle and 
the $\tilde{e} - \tilde{\mu}$ mass splitting from lepton
flavor changing processes. For example, $\mu \rightarrow e \gamma$ 
gives an important constraint. For degenerate left-handed sleptons, 
and with the LSP $(\chi^0 _1)$ approximately bino-like $(\tilde{B}^0)$, 
the constraint on $\sin 2 \theta _R$ and the mass splitting $\Delta m$ 
between 
the right-handed sleptons is 
approximately  
\begin{equation}
\sin 2\theta_R (\Delta m) /m \stackrel{<}{\sim} 
0.01 \times \sqrt{\frac{BR( \mu \rightarrow e \gamma)}
{4.9 \times 10^{-11}}}.
\label{muegamma}
\end{equation}
(A more proper 
formula is given in Section \ref{LHCpoint}).

Suppose a real 
selectron is produced in the basis of Eqn.(\ref{basis2}) (say
in association with an electron). Since $\tilde{e} \; (\alpha =1)$
is not a mass eigenstate, there is a probability that as it
propagates it will convert to a $\tilde{\mu} \; (\alpha =2)$
and hence decay into a $\mu$ \cite{nima1,nima2}:
\begin{equation}
P(\tilde{e}_{\alpha = 1} \rightarrow \mu \chi_1^0 ) = 2 \sin ^2 \theta 
\cos ^2 \theta \; x,
\label{x}
\label{osc}
\end{equation}
where $x = (\Delta m)^2 / \left( (\Delta m)^2 + \Gamma ^2 \right)$ is
the quantum
interference factor and 
assuming $BR ( \tilde{l} 
\rightarrow l \chi_1^0 ) = 1$. 
Here $\Gamma$ is the decay width of the slepton.
Note that for $\Delta m \stackrel{>}{\sim} \Gamma$ the interference effect can 
be neglected so that $x \sim 1$. In this case the oscillation 
probability can be large.
Typically, $\Gamma \sim \alpha_{em} m \sim 0.01 m$ so that $x \sim 1$
if $(\Delta m)/m \stackrel{>}{\sim} 0.01.$  
This is close to the upper bound from the $\mu \rightarrow e \gamma$ limit, 
so there could be a suppression due to either $x$ or $\sin \theta$ 
\cite{nima1,nima2}.
It is possible, however, that for a specific SUSY spectrum the decay 
width could be much smaller than this naive estimate, allowing for a larger 
range of $\Delta m /m$ consistent with the rare decay limit (even
for large 
mixing angles) {\em and}
$x \sim 1$ so that
the oscillation
signal is not suppressed \footnote{In fact, this occurs at the LHC Point 
discussed
in Section \ref{LHCpoint}.}. 

Similarly, a neutralino can decay into
$e^+ \; \mu^-$ or $e^- \; \mu^+$ through an intermediate slepton:
\begin{equation}
\chi_2^0 \rightarrow \tilde{l}^+ l^- \hbox{, } \tilde{l}^- l^+
 \rightarrow l^+ l^- \chi_1^0.
\label{chi2a}
\end{equation}
Using Eqn.(\ref{osc}) the rate for a flavor violating decay is 
\begin{eqnarray}
BR ( \chi_2^0 \rightarrow e^+ \mu^- \chi_1^0)  
&=& 2 \;
\sin ^2 \theta \cos ^2 \theta \; x 
 \times  BR ( \chi_2^0 \rightarrow 
\tilde{e}^- e^+, \tilde{\mu}^+ \mu^-).
\label{chi2bi}
\end{eqnarray}
Here to simplify notation $BR(\chi_2^0 \rightarrow      
\tilde{e}^- e^+, \tilde{\mu}^+ \mu^-) \equiv
BR(\chi_2^0 \rightarrow
\tilde{e}^- e^+)+BR(\chi_2^0 \rightarrow\tilde{\mu}^+ \mu^-)$. This 
notation will be used throughout the paper. 
Also, the  
BR on the right-side of Eqn.(\ref{chi2bi}) is in the absence of any mixing.
In the case of interest here of small mass splittings, $\Delta m \ll m$,
the neutralino decay rate into selectrons or smuons are equal 
in the absence of any mixing. Next, in the absence of mixing,
\begin{equation}
BR(\chi_2^0 \rightarrow e^+ e^- \chi^0 _1) =2 \; 
BR(\chi_2^0 \rightarrow \tilde{e}^+ e^-) .
\end{equation}
The factor of two occurs since 
$\chi_2^0$ may decay to $\tilde{e}$'s of both charges. This 
result and Eqn.(\ref{chi2bi}) relates the flavor-violating  
and flavor-conserving decays:
\begin{eqnarray}
BR ( \chi_2^0 \rightarrow e^+ \mu^- \chi_1^0) =& 2 \; 
\sin ^2 \theta \cos ^2 \theta \; x 
\times BR ( \chi_2^0 \rightarrow l^+ l^- \chi^0 _1), 
\label{chi2b}
\end{eqnarray}
where the BR on the right-side of the above equation is in the
absence of mixing.
Here $l$ is either $e$ or $\mu$.
This result applies for $\chi_2^0$ decays to real sleptons, {\it i.e.,}
for $m_{\chi_2^0} > m_{\tilde{l}}$. 
For $m_{\chi_2^0} < m_{\tilde{l}}$, there is an 
additional suppression of $(\Delta m)/m$ in 
the decay amplitude due to the supersymmetric analog of the
Glashow-Iliopoulos-Maiani (GIM)
cancellation as in the case of $\mu \rightarrow e \gamma$,
resulting in negligible $e \mu$ signal. So an observable $e \mu $
signal requires the production of real sleptons \footnote{Alignment
models with $\Delta m \sim m$ are not considered here since $\sin \theta 
\sim O(10^{-2})$.}.   

\section{Slepton Production by Drell-Yan Process}
\label{dy}
One way to produce sleptons at a hadron collider is through the
Drell-Yan process:
\begin{equation}
p \; p (\hbox{or} \; \bar{p}) 
\stackrel{\gamma,Z}{\rightarrow} \tilde{l}^{\ast} \tilde{l} \rightarrow
l^+ l^- \chi_1^0 \chi^0_1.
\end{equation}
Thus the production of sleptons is identified by  
events with no jets, 
2 hard isolated leptons and $ \not \! p_T$, assuming that 
$\chi^0 _1$ is stable or decays outside the detector.  
These events will be referred to as ``flavor conserving''
dilepton events. 

There is a SM background to the signal
from $W^+ W^-$ and $\bar{t} t$ production. These backgrounds are known,
in principle. In \cite{baer} a set of 
kinematic cuts on the leptons, as well as a jet--veto, are found which 
sufficiently reduce these backgrounds. 
These cuts reduce the signal as well -- of course, the reduction is much more
for the background. 

There is also a
SUSY background from $p p \rightarrow
\chi ^ + \chi ^ - 
\rightarrow W^+ W^- \chi _1 ^0 \chi _1 ^0$. 
This background
depends
on the model--dependent $\chi^+ \chi^-$ production cross section.
But, for supergravity motivated parameter choices with
$m_{\tilde{q}} \approx m_{\tilde{g}}$, this background can 
be sufficiently reduced by using the same cuts used to remove the SM 
background \cite{baer}. 
For example, from the analysis of \cite{baer} (see Table III of \cite{baer})
with $10$ (fb)$^{-1}$ and for a 
slepton mass $\sim 100$ GeV 
there are $\sim 20$
signal events with no background events remaining after the cuts. 

Actually, a clever method \cite{krasnikov1}
for detecting the sleptons is to form the
asymmetry $A_F=N(e^+e^- + \mu^+ \mu^-)-N(e^+\mu^- + e^-\mu ^+)$.
The background does not contribute to $A_F$, so a non-zero
value would provide evidence for slepton production.

In the lepton flavor mixing case the pair production of 
sleptons will produce 
$e \mu$ events with $ \not \! p_T$ -- these 
events will be referred to as ``flavor
violating'' dilepton events. 
The background to this signal is from the same sources as for the
flavor conserving dilepton signal (with the 
same rate) as well as from $\tilde{\tau} \tilde{\tau}^{\ast}$ production
followed by leptonic decays of $\tau$s. 

The
detection of SUSY lepton flavor violation 
using the above
flavor violating dilepton events for the CMS detector at the LHC
was studied in references \cite{krasnikov2,krasnikov1} for
the case of maximal mixing $(\theta = \pi/4)$. 
With the mixing angle being maximal, the flavor violating dilepton
signal rate is high; see Eqn.(\ref{osc}) (assuming $x \sim 1$). 
In fact, the number of flavor conserving and flavor
violating events from slepton production in this case are the same and 
each is equal
to one half the signal in the zero mixing case so that
$A_{F} \approx 0$ (unlike the case of zero or non-maximal
mixing). In the case where
the production cross--sections for
staus and the lightest charginos are comparable to that of the sleptons,
the production rate for the
SUSY background to $e \mu$ events
is $\sim 4 \%$ of the total
flavor conserving signal (in the absence of
mixing). \footnote{Here, it is assumed that $BR \left( \chi^+
\rightarrow W^+ \chi_1^0 \right) \approx 100 \%$ so that
the leptonic BRs of $\chi^+$ are the same as for $W$. 
If the left-handed sleptons
are lighter than $\chi^+$, then the leptonic BRs of $\chi^+$ may be enhanced
substantially, in turn increasing the SUSY background.}
Thus, the chargino and stau backgrounds
are
much smaller. 
The high signal and low SUSY background rate (compared to the signal)
for maximal mixing enables detection of a $
5\sigma$ flavor violating signal
for sleptons masses up to 250 GeV and LSP masses $m_{\chi^0 _1} < 
0.4 m_{\tilde{e}_R}$ with an integrated luminosity of
$100$ fb$^{-1}$. 

There are some objections to the generality 
of this result, though. A more general spectrum could 
result in a larger chargino or stau background. For example, 
there is no reason to expect the
chargino production cross--section to be related to the slepton
production cross-section. However, 
as mentioned above,
the kinematics of slepton
production and decay are different enough from that of
the chargino background
that an appropriate set of kinematic cuts
could distinguish the two, at least for supergravity motivated
parameter choices with comparable
squark and gluino masses \cite{baer}.
Next, the stau background
is sensitive to the stau mass, which is likely to differ from the
selectron and smuon masses
\footnote{The rare decays
$\tau \rightarrow e \gamma$, $\tau \rightarrow \mu \gamma$ and
$\mu \rightarrow e \gamma$ allow for $O(1)$ splitting between the
third and first two generation scalars for $CKM-$like
mixing angles.}. 
The stau background has softer leptons, so
a cut on the $p_T$ of the leptons may help distinguish this background from
the signal. The success of this
may require large statistics and knowledge of the stau
production cross--section.
Thus, in general, the SUSY background may not be small.

Next, detection of flavor violation for smaller mixing angles is discussed.
Since the signal is
$\propto \sin^2 \theta$, it is significantly
smaller for say Cabibbo-like mixing angles. In this case, it is crucial
to know the SUSY background more precisely
since it is comparable to the
signal (assuming similar cross sections for sleptons and charginos).
While the quantity $A_F$ ($> 
0$ for non-maximal mixing)
is, up to statistical fluctuations,   
background--free as far as slepton {\em detection} is concerned, it  
is not useful for providing evidence 
for slepton {\em flavor violation} since 
the chargino background  
would need to be determined first. 
This is 
because the deviations in the values of $A_F$ and $N(e \mu)$
from the SM  
for a non-zero mixing angle could
be reproduced, in the case of zero mixing angle, with  
a lower slepton production cross--section 
and a higher chargino production cross--section.   

Even if the SUSY background can be reduced sufficiently by an 
appropriate set of cuts,
since the signal is suppressed by the small mixing angle (there will
also be a reduction of the signal due to these cuts), 
it may not possible to probe Cabibbo-like mixing angles. For example,
in the case of no mixing,    
Table 4 of reference 
\cite{krasnikov1} gives 195 dilepton signal events
for the set of cuts labeled 1 
with  
$L=10$fb$^{-1}$ and a slepton mass of $100$
GeV. The number of signal events in the case of
mixing for $L=100$fb$^{-1}$
is then  
$1950 \times  
2\times \sin^2 \theta \cos ^2 \theta$ (assuming $x \sim 1$). 
The SM background from $WW$ production is 9920 
for the same set of cuts. Thus a $5\sigma$ signal (requiring 
$S/\sqrt{B} >5$) is possible only for $\sin \theta \stackrel{>}{\sim}
0.4$. 
Since this signal was obtained for a 24 GeV LSP, only larger 
angles will be probed for larger LSP masses (since the leptons
will be softer in that case). For sleptons heavier 
than 100 GeV the prospects for detecting small mixing 
angles are clearly worse.
 
Thus, in the situation where
the SUSY background is {\em known} to be small, {\em e.g.} if an 
appropriate set of cuts for a more general spectrum 
can separate the chargino background 
from the signal, then the 
flavor violating dilepton events from Drell--Yan production of 
sleptons is a promising signal for the 
detection of flavor violation in the case of {\em large} mixing angles.
Otherwise, it is important to look for 
other discovery channels for slepton flavor violation.

\section{Slepton Production in Cascade Decays}
The other way to produce sleptons is through the cascade decays
of gluinos and squarks. At the LHC,
the production cross sections of squarks and gluinos
are much
larger than the Drell-Yan production of sleptons, 
neutralinos, and charginos.
So, a larger production of sleptons (if they are light)
is expected in the cascade decays than
from direct Drell-Yan production.
In a generic SUSY event, the production of 
two real (or virtual from gluino
decay) squarks will be followed by their cascade decays ultimately
to the LSP through intermediate electroweak sparticles 
(sleptons, charginos, neutralinos).
Assuming for simplicity
that the spectrum is gaugino-like,
$i.e.$, 
$\chi_2^0 \approx \tilde{W}_3$, $\chi_1^+ \approx \tilde{W}^+$ and
$\chi_1^0 \approx \tilde{B}$, the following squark decays
are obtained:
\begin{eqnarray}
BR ( \tilde{q}_R \rightarrow q \chi_1^0 ) & \approx & 1 ,\nonumber \\
BR ( \tilde{q}_L \rightarrow q \chi_2^0 ) & \approx &
\frac{1}{3}, \nonumber \\
BR ( \tilde{q}_L \rightarrow q^{\prime} \chi_1^{+,-}) & \approx &
\frac{2}{3}.
\end{eqnarray}
Thus, a typical SUSY event is:
\begin{eqnarray}
p p & \rightarrow & \tilde{g} \tilde{g}, \; \tilde{g} \tilde{q} 
\rightarrow
\tilde{q} \tilde{q} \nonumber \\
 & \rightarrow & \chi_{EW} \chi'_{EW} \; + \; X, 
\end{eqnarray}
with $\chi_{EW}$, $\chi'_{EW}$ one of $\chi^0_{1,2}$, $\chi^{+,-}_1$.

\subsection{Dilepton Events}
\label{cascade2l}
If 
one of the squarks decays to $\chi_2^0$ followed
by the decay of $\chi_2^0$ to a slepton (if
$BR( \chi_2^0 \rightarrow \tilde{l} l)$ is significant)
a large number of
$e \mu$ events in the presence of
lepton flavor mixing (see Eqns.(\ref{chi2a}) and (\ref{chi2b})) 
is obtained.
These events also have 
at least $2$ high $p_T$ jets and large
$ \not \! p_T$. 

There is no background from $W^+W^-$ production since this background
contains no hard jets (assuming jet detection is good).
There is a background from
$t \bar{t}$ production followed by leptonic decays
of the $W$'s from the top quarks. This can be reduced by 
rejecting events with $b$-jets
or using a high
$\not \! p_T$ cut.

There is a SUSY background from 
the decays of both
squarks to charginos,  
followed by chargino decays 
to $W^+$, $W^-$ or $\tilde{l}$, $\tilde{l}^{\ast}$.  This background is
distinguishable from the signal though.
The invariant mass distribution of the 2 leptons from the $\chi_2^0$ decay
has a sharp edge (which is a 
function of the neutralino and slepton masses) \cite{snowmass,ian}
unlike the case of
the $2$ leptons from $\chi^+ \chi^-$ decays.
Also, 
the angle between the 2 leptons from the decay of $\chi_2^0$ is likely to
be smaller than in the case of 2 leptons from $\chi^+$ and $\chi^-$.
Such kinematic cuts on the invariant mass of the dileptons and
the angle between them 
easily reduce the number of background events
sufficiently if we are interested in
detecting flavor {\it conserving}  dileptons from
$\chi_2^0$ decays. 

But, in the case of the flavor {\it violating} dilepton events,
(as in Section \ref{dy}) since the signal 
is suppressed by the mixing angle (while the background is the same),
the number of background events that survive (relative to the signal)
after cuts
depends crucially on 
the model--dependent cross sections for producing
$\chi^+ \chi^-$ vs. $\chi_2^0$ 
 \footnote{
For example, the ratio of the number of events with
$\chi^+ \chi^-$ to those with
(at least) $1 \; \chi_2^0$ 
is larger for 
$s$-channel $\tilde{q} \tilde{q}^{\ast}$
production than for gluino pair production which is seen as follows.
For the $\tilde{g} \tilde{g}$ case, the probability of getting
$2 \; \tilde{q}_L$ is $1/4$ compared to a probability of 
$3/4$ for getting at least one $\tilde{q}_L$
whereas for $s$-channel $\tilde{q} \tilde{q}^{\ast}$ production the
probabilities are the same. {\em Same} sign
chargino events are also obtained from $\tilde{g} \tilde{g}$ 
production whereas $s$-channel $\tilde{q} \tilde{q}^{\ast}$ production
can give only opposite-sign chargino pairs. 
Thus, if the $s$-channel $\tilde{q} \tilde{q}^{\ast}$ production
is larger, the number of $\chi^+ \chi^-$ events
relative to $\chi _2 ^0$ events increases.}. So in general 
it is difficult to be sure that the cuts have reduced the 
background sufficiently. 
%
\footnote{
There is also a 
SUSY background from
$\chi_2^0$ decays to $\tilde{\tau} \tau$
followed by leptonic decays of the $\tau$'s. A
cut on the dilepton invariant
mass
can reduce this: the leptons from the $\tau$ decays are softer
than those from the $\tilde{e} / \tilde{\mu} / \chi_2^0$
decays and so have a smaller invariant mass.
But, since, in general,
$BR \left( \chi_2^0 \rightarrow \tilde{\tau} \tau \right)$
is {\em not} related to $BR \left( \chi_2^0 \rightarrow \tilde{e} e \right)$,
as for the chargino background, we cannot be sure that the $\tilde{\tau}
\tau$ background has been sufficiently reduced (by the cuts)
since this
background is unknown.}         

In the circumstance that $\chi^+ \chi^-$ are 
dominantly produced from $\tilde{g} \tilde{g}$  
cascade decays,
the $\chi^+ \chi^-$ flavor violating 
background can be estimated as follows.
An equal number of like-sign and 
unlike-sign chargino pairs are expected 
since $\tilde{g}$ is a Majorana particle.
The like-sign chargino pairs produce like-sign dileptons so that
the opposite-sign chargino $e \mu$ background 
can be estimated from the number of like-sign $ee$ and
$\mu \mu$ events. 
Unfortunately, in the more general case 
the $\chi^+ \chi^+$ and $\chi^+ \chi^-$
production cross sections are 
not related since the chargino pairs do not
always come from gluino pair decays. 
For example,
$p p \rightarrow \tilde{q}_L \tilde{q}^{\ast}_L$ can lead to $\chi^+ 
\chi^-$, but
not to $\chi^+ \chi^+$. 

It might be possible to estimate the $\chi^+ \chi^-$
background by analysing the (observed) (signal $+$ background)
distribution of the invariant mass of
the flavor violating dileptons \cite{paige}. As
mentioned earlier, the dilepton invariant mass distribution for
$\chi_2^0$ decay has a
sharp edge unlike the case of
the background. The position of this
edge (denoted by $M_{ll}$)
can be easily found by looking at the distribution
of the invariant mass of flavor
{\em conserving} dileptons (where the
$\chi^+ \chi^-$
background is very small) \cite{snowmass, ian}. 
The existence of 
an 
edge in the (observed) {\em opposite}--flavor dilepton distribution 
{\em at  $M_{ll}$} would then be 
indication of flavor violation. However, since the flavor violating dilepton
signal is suppressed by (small) mixing relative to the 
flavor conserving dilepton signal (whereas the $\chi^+ \chi^-$
background is the same for both kinds of dileptons), the edge at
$M_{ll}$ in the opposite flavor
dilepton case might not be as sharp as for the same flavor dilepton 
case -- this depends
on the model-dependent cross sections for producing $\chi^+ \chi^-$ vs.
$\chi_2^0$. 

Next, in the distribution of the invariant mass of
the flavor {\em violating} dileptons, the events beyond 
$M_{ll}$ 
(this value can be obtained from the same flavor dilepton distribution if
the edge is not so sharp in the opposite--flavor 
dilepton distribution)
are mostly from the $\chi^+ \chi^-$
background \cite{paige}.
Extrapolating (assuming say a flat distribution for the
$\chi^+ \chi^-$
background) from the data in this region,
the
$\chi^+ \chi^-$
background in the region
with invariant mass less than $M_{ll}$
can be estimated. An excess of $e \mu$ events
(with invariant mass between zero and $M_{ll}$)
over this estimate
will be a signal for flavor violation.
\footnote{The invariant mass
of the leptons from the $\tilde{\tau} \tau$ decays (from $\chi_2^0$) is less
than $M_{ll}$ and so this background cannot be estimated this way.}
This extrapolation may
not be reliable
for invariant masses much smaller than $M_{ll}$
since the
distribution of the $\chi^+ \chi^-$
background in this region is not known.
A detailed simulation is required
to know this distribution (it is known only
that it does not have an edge at $M_{ll}$).
Near $M_{ll}$ the extrapolation should
be more
reliable and that is the region where the signal is peaked (since
the flavor violating dileptons from $\chi_2^0$ decay also have
a sharp edge at $M_{ll}$). An excess in this region
(rather than the whole region between zero mass and $M_{ll}$)
might thus
be a better signal for flavor violation \cite{paige} -- as
mentioned earlier, the
distribution will have a edge (or a ``step'') at $M_{ll}$. 
Also, the $\tilde{\tau} \tau$ background in the region near $M_{ll}$
is negligible since the leptons from these decays are softer \cite{paige}.
However,
statistics
are larger if the region from zero mass to $M_{ll}$ is used.


The chargino background can also be eliminated in 
considering a flavor violating {\em and} $CP$ violating 
dilepton signal \cite{nima2}.  The presence of non--trivial phases in the 
slepton mixing matrix $W$ breaks $CP$, and results in a 
non-vanishing asymmetry:
$N( e^+ \mu^-  -
e^- \mu^+) \neq 0$.   
In this case, the $\chi^+ \chi^-$ background
is not important since it is CP symmetric.

To summarize, if the  
{\em number} 
of $e \mu$
events (that pass certain
cuts) from either Drell-Yan or cascade production is used 
to detect flavor violation, the SUSY
background from $\chi^+ \chi^-$ pairs (which passes the same cuts)
is difficult to estimate, in general, and may be too large.
The possibility of using the {\em observed}
opposite--flavor dilepton mass {\em
distribution} (in the case of cascade decays) to estimate the
chargino background 
is interesting, though, and warrants further study \cite{paige}.

\subsection{Events 
with  
$4$ leptons}
\label{cascade4l}
A dramatic flavor violating signal is obtained through the 
pair production of {\it two} $\chi^0 _2$s, followed by the 
decays of {\em both} $\chi_2^0$s to slepton and lepton pairs.
Such an event contains 4 leptons and occurs 
if both squarks in a SUSY event decay into $\chi_2^0$.
If one of the $\chi_2^0$s has a
flavor violating decay: $\chi_2^0 \rightarrow \tilde{l} l \rightarrow
e \mu$, then events containing  
$3 e \; 1 \mu$, or $3 \mu \; 1 e$ will be produced. A typical 
decay chain is then:
\begin{eqnarray}
\tilde{q}_L \tilde{q}' _L 
&\rightarrow& \chi^0 _2 q + \chi^0 _2 q' \nonumber \\
\chi^0_2  &\rightarrow& \tilde{l} l \rightarrow 
e^+ e^- \chi^0 _1 \nonumber \\
\chi^0_2 &\rightarrow&  \tilde{l}' l' \rightarrow 
\mu ^+ e^- \chi^0 _1 \;.
\end{eqnarray}
These events are identified by 4 isolated leptons (with the 3+1 flavor
structure), at least
$2$ high
$p_T$ jets, $ \not \! p_T$, and concentrating on only those 
events produced from the decays of first two generation squarks, 
no $b-$jets. These events will be referred
to as  ``flavor violating'' 4 lepton events. 
The absence of $b-$jets is important in distinguishing the signal 
from other SUSY and SM backgrounds (see below). 

The backgrounds to these events arise from both SM and SUSY 
sources.  

The dominant SM 
background occurs from $t \bar{t}$ production 
with semileptonic decays of 
the $b$s (or $t \bar{t} \gamma$
production with $2$ leptons
from $\gamma$)
and leptonic decays of the $W$s. In this case, however,  
the leptons from $b$ decays
will not be isolated (or
the invariant mass of $2$ of the leptons will be zero
in the case of $t \bar{t} \gamma$). Also, 
these events have $2$ $b$ quarks and
can be rejected using $b$-jet veto. Double gauge 
boson production can give 4 lepton events, but none of these 
events have the 3+1 flavor structure. 
Triple gauge boson production ($WWZ$ or $WW\gamma$) 
can give events with
$4$ leptons and the correct flavor asymmetry, 
but some initial state gluon radiation is 
needed to give the
$2$ hard jets. The production cross-section for such events 
is small. Also, 
events of this kind can also be rejected since the invariant mass
of $2$ of the leptons will either be zero 
or $m_Z$. 

One important obstacle in identifying flavor-violating {\it dilepton} 
events was the potentially large background from $\chi^+ \chi^-$ 
production. In the $4$ lepton signal, however, there is no $\chi^+ \chi^-$ 
background 
from the squark decays since this gives only $2$ leptons. 

The weak decay  
$\tilde{q} \rightarrow W \tilde{q}'$, if kinematically allowed, 
can lead to a possible background. For example, the process 
\begin{eqnarray}
\tilde{q}_L \tilde{q}'_L &\rightarrow& W^- 
\tilde{q}'' _L + \chi^+ _1 q 
\nonumber \\
W^- &\rightarrow& e^- \bar{\nu} \nonumber \\
\tilde{q}'' _L &\rightarrow& \chi^0_2 q '' 
\rightarrow \mu^{+} \mu^{-} + \cdots 
\nonumber \\
\chi^+ _1 &\rightarrow& \mu^+ \nu \chi^0_1 
\end{eqnarray}
is a potential background. For the first two generation squarks, however, 
the decay $\tilde{q} \rightarrow W \tilde{q}'$ is kinematically forbidden. 
This is because the   
mass splitting in an electroweak doublet occurs from the electroweak 
$D-$terms 
and is less than 
$ m^2_W /m_{\tilde{q}} < m_W$. This process is allowed for 
the top and bottom squarks, but such an event contains 2 $b$-jets 
and this background can be reduced with a $b$-jet veto. 

There is a 
SUSY background to the flavor violating
$4$ lepton events
from production of heavier neutralinos or chargino
in the cascade decays of
squarks.
For example,
\begin{eqnarray}
\tilde{q}_L \tilde{q} & \rightarrow & \chi^0_3 \chi_2^0 + \cdots,
\nonumber \\
\chi_3^0 \rightarrow W^+ \chi^-  & \rightarrow & e \mu + \cdots, \nonumber \\
\chi_2^0 \rightarrow \tilde{l} l & \rightarrow & ee \;
(\hbox{or} \; \mu \mu) \chi^0_1. 
\end{eqnarray}
This background is small in the so-called 
gaugino-like region. In this region  
there is very little gaugino-Higgsino mixing. Then,
the heavier chargino and the two heaviest
neutralinos are dominantly Higgsinos and the two lightest neutralinos
and the lighter chargino are mainly gauginos; this turns out to be
typical of the
SUSY parameter space still allowed by experimental data. 
Thus, the decays of the  
first two generation squarks into the heavier 
neutralinos or chargino 
are highly suppressed by the first two generation
Yukawa couplings, 
small gaugino-Higgsino
mixing, 
and also by phase space.

Another potentially large background can also occur from the production of 
the {\em heavier} sleptons (say, the left-handed) and/or sneutrinos. 
 Sleptons
can decay 
to $\chi^0 _2 l$ and $\tilde{\nu}$ to  
$\chi^{\pm}_1 l$ if kinematically 
allowed.  
If the neutralino and chargino decay to leptons, then
this decay chain can give $3$ (or $2$) leptons. With $1$ (or $2$) leptons
from another decay of this kind (or some other decay chain), this
can mimic
the flavor violating $4$-lepton signal.
If 
the left-handed sleptons are paired produced through the Drell-Yan 
mechanism, then these events do not contain any hard jets and may be rejected. 
Thus, the only source for a background from heavier sleptons is 
their 
production in the decays 
of gluinos and squarks. Such a decay does not occur directly, but only 
through the decays of gluinos and squarks to the {\em heavier} neutralino 
and chargino. The {\em heavier} neutralinos and chargino can then decay 
to the left-handed sleptons. As argued in the previous paragraph though, 
in the gaugino-like region, the heavier neutralinos/chargino are dominantly
Higgsinos so that their decays to the sleptons are suppressed
by the lepton Yukawa couplings and small gaugino/Higgsino mass 
mixing angles. So 
this background is negligible.    

However, top squarks 
(and bottom squarks
for large $\tan \beta$) will have significant decay branching fractions
into heavier neutralinos or chargino even if they are
purely Higgsinos since the third generation Yukawa couplings (and hence
couplings of the squarks to Higgsinos) are large. Further, as mentioned 
earlier,  
$W$s may be produced in the direct decay of stops or sbottoms.
Also,
top {\em quarks} from stop or sbottom decays produce $W$s. Both of these 
processes give  
additional isolated leptons. This leads to a potential background 
even if stops or sbottoms
decay only to the lighter chargino and neutralinos. For example,
the following decay chain is a possible background:
\begin{eqnarray}
\tilde{t} \tilde{t}^{\ast} & \rightarrow & b \chi^-  +
t \chi_2^0, \nonumber \\
t & \rightarrow & W^+ b \rightarrow e^+ \; b + \cdots, 
\nonumber \\
\chi^- & \rightarrow & W^- \chi^0_1  \rightarrow \mu^- + \cdots,
\nonumber \\
\chi_2^0 & \rightarrow & e^+ e^- \chi_1^0.
\label{stopdecay}
\end{eqnarray}
These backgrounds to flavor violating
$4$ lepton events can be reduced 
by  
rejecting any $4$ lepton event that contains at least one $1$ $b$-jet.
Note that the top or bottom squark background   
has at least $2$ $b$ quarks.  
The efficiency for rejecting this 
background is discussed in a later section where a specific 
spectrum is considered. 

There is also an important 
SUSY background from decays of taus and staus 
produced from the decays of two $\chi _2^0$s. That is,
\begin{eqnarray}
\chi _2^0 & \rightarrow & \tilde{\tau} \tau  \rightarrow e \mu 
\chi^0_1 + ...,
\nonumber \\ 
\chi _2^0 & \rightarrow & \tilde{l} l \rightarrow  ee \; (\hbox{or} 
\; \mu \mu) \chi^0_1.
\end{eqnarray}
This background can be estimated/measured as follows.
In the above decay chain, if one $\tau$ decays hadronically instead of
leptonically, the result is  
$3 e \; 1 \tau$-jet events. 
If a lower bound on the $\tau$-jet detection efficiency is known, 
an upper bound on the number of 
$3 e \; 1 \mu$ events coming from $\tau$ decays is obtained 
by using the number
of $3 e \; 1 \tau$-jet events. An 
excess of $3 e \; 
1 \mu$ events
over this background is a signature of lepton flavor violation. 

Lastly, the following $\chi_2^0$
decay chains can also give flavor violating dileptons:
\begin{eqnarray}
\chi_2^0 & \rightarrow & h \; (\hbox{or}) \; Z \; \chi_1^0 \nonumber \\
h \; (\hbox{or}) \; Z & \rightarrow & \tau \tau \rightarrow e \mu. 
\end{eqnarray}
In combination with another $\chi_2^0$ decay to $ee$ or $\mu \mu$,
these decay chains can give flavor violating $4$-lepton events.
In the gaugino region, the decay $\chi_2^0 \rightarrow Z \chi_1^0$ is
suppressed since there is no vertex with $Z$ and $2$ neutral gauginos. 
In any case, an {\em effective} $BR \left( \chi_2^0 \rightarrow
\tau \tau \right)$ can be defined to include these two decay chains
in addition to the $\chi_2^0 \rightarrow \tilde{\tau} \tau$ decay.
It will be shown in section \ref{LHCpoint} that this 
(in general unknown) BR does not affect
the estimate of the (effective) $\tau$ background
obtained by using the $3 e \; 1 \tau$-jet events.  

\subsubsection{A quick estimate of number of $4$ lepton events}
\label{quick}
A typical value for the total SUSY production cross section 
(gluinos and squarks) at the LHC is:
\begin{equation}
\sigma_{SUSY} \sim 10 \;
\frac{\pi \alpha_S^2}{\hat{s}} \sim 100 \; \hbox{pb} 
\end{equation}
with $\sqrt{\hat{s}} \sim 1$ TeV, $\alpha_S \sim 0.1$ and summed over
colors and generations (the factor of 10).
Assuming that the probability to get a $\tilde{q}_L$ 
is $1/2$ and $BR ( \tilde{q}_L \rightarrow
\chi_2^0 \; q) = 1/3$, this gives 
\begin{equation}
\sigma_{\chi^0_2 \chi^0_2} \sim \sigma_{SUSY} \left( \frac{1}{2} \right)^2 
\left( \frac{1}{3} \right)^2 \sim 3 \; \hbox{pb}. 
\end{equation}
If $BR ( \chi _2^0 \rightarrow \chi^0 _1 l^+ l^-) 
\sim 0.16$ (for each of $l = e$, $\mu$) and 
for $\sim$ one year of running at low
luminosity which gives an integrated luminosity of 
$L \sim 10$ (fb)$^{-1}$, 
the expected number of events is :
\begin{eqnarray}
N \left( \chi_2^0 \chi_2^0 \right) &&  \sim 30,000, \nonumber \\
N \left( 4 \; l \; \hbox{where} \; l = e, \mu 
\right)&=&(2 \times 0.16 )^2 
N\left( \chi_2^0 \chi_2^0 \right)  \sim \; 3300, \nonumber \\
N \left( 3 l + l'  \right) &=&
4 \sin^2 \theta \cos^2 \theta \; x \; N(4 l) \sim \; 550,
\end{eqnarray}
for $\sin \theta \approx 0.2$ and $x\sim 1$.
To be clear, 
\begin{equation}
S_{FV} \equiv N(3l+l')=\left(N(e^+ \mu^- \mu^+ \mu^-)+
(+ \leftrightarrow -)\right)+(\mu \leftrightarrow e),
\end{equation}
and
\begin{equation}
N(4l)=N(e^+ e^- e^+ e^-)+N(e^+ e^- \mu^+  \mu^-)
 +N(\mu^+ \mu^-  \mu^+ \mu^-)
+ S_{FV}.
\label{all4l}
\end{equation}
In the next section this definition for $N(4l)$ is trivially extended 
to include 
leptons produced by the decay of $\tau$s. 
Thus, typically, a large number of $4$ lepton
flavor violating events is expected from the cascade decays of squarks 
\footnote{{\em Both} $\chi_2^0$s decaying to flavor violating dileptons
gives $(e^+ \mu^-)(e^+ \mu^-)$ and $(e^+ \mu^-)(\mu^+ e^-)$
events. The latter cannot be distinguished from the events where
one $\chi_2^0$ decays to $e^+e^-$ and the other to $\mu^+ \mu^-$.
The former events can also be used as a signal of flavor violation, but
the number of these events is expected to be very small since they require
both $\chi_2^0$s to decay into flavor violating dileptons.
For simplicity these events were not included in Eqn.(\ref{all4l}).}.

\subsubsection{Detailed estimates at Point $5$ (Point $A$) of LHC studies} 
\label{LHCpoint}

One Point of the LHC supersymmetry studies \cite{snowmass,ian} contains 
a spectrum that 
is favorable for the detection of a flavor-violating 4 lepton signal. 
The minimal supergravity  input parameters for this point are:
\begin{eqnarray}
 m_0 = 100 \hbox{ GeV,} \; & M_{1/2} = 300 \hbox{ GeV, } 
\;& A_0 = 300 \hbox{ GeV,}
\nonumber \\
\tan \beta = 2.1 \;,& \; \hbox{sgn}(\mu) = + \;, & \; 
m_{top} = 170 \hbox{ GeV.} 
\end{eqnarray} 
Renormalization group evolution of these input parameters to the weak scale 
results in a mass spectrum which is given in Table \ref{spectrum}.
Note that $m_{\chi_2^0} \approx 230 \;
\hbox{GeV} \; > m_{\tilde{l}_R} \approx 160$ GeV so that the decay of 
$\chi^0 _2$ into real sleptons is allowed.  
\begin{table}
\begin{center}
\begin{tabular}{llllllll} \hline
$\tilde{g}$ & 770 & $\tilde{q}_L$ &685 & $\tilde{q}_R$ & 660 &$h$ & 100 
\\ \hline
$\tilde{t}_1$ & 500 & $\tilde{t}_2$ &715 & $\tilde{b}_1$ &635 
& $\tilde{b}_2$ &660 \\ \hline
$\tilde{l}_L $ &240 & $\tilde{l}_R$ & 160 & $\chi^{\pm} _1$ &230 &
$\chi^{\pm} _2$ & 500 \\ \hline
$\chi^0 _1$ & 120 & $\chi^0 _2$ & 230 & 
 $\chi^0 _3$ &480 & $\chi^0 _4$ & 505 \\ \hline
\end{tabular}
\end{center}
\caption{Mass spectrum in GeV at LHC Point \protect\cite{snowmass,ian}.
Here $\tilde{q}= \tilde{u}$, $\tilde{d}$, $\tilde{c}$, $\tilde{s}$, and 
$\tilde{l}= \tilde{e}$, $\tilde{\mu}$, $\tilde{\tau}$.}
\label{spectrum}
\end{table}

\begin{table}
\begin{center}
\begin{tabular}{cccccc} \hline
$\tilde{g} \tilde{g}$ & 1750 & $\tilde{g} \tilde{q} \;, \; 
\tilde{g} \tilde{q}^{\ast}$ & 8300  &
$\tilde{q} \tilde{q}^{\ast}$ & 2380 \\ \hline 
$\tilde{q} \tilde{q}'$ & 2820 & 
 $\tilde{b} \tilde{b}^{\ast}$ & 300
& $\tilde{t} \tilde{t}^{\ast}$ & 700 \\ \hline
\end{tabular}
\end{center}
\caption{The production cross-sections in fb for different SUSY particles 
at the LHC Point \protect\cite{snowmass,ian}. Here all flavors 
$\tilde{q}_H=\tilde{u}$, $\tilde{d}$, $\tilde{c}$,
$\tilde{s}$ 
and $H=L$, $R$ are summed over.}
\label{cross-section}
\end{table}
The production cross-section for SUSY particles is presented in 
Table \ref{cross-section}, and is dominated by $\tilde{g} \tilde{q}$ 
production.  
In total   
$\sigma_{SUSY} \approx 16$ pb. To estimate the number of 
signal and background events, the branching fractions of the sparticles 
are needed. These are given in Table \ref{br}. Note that at this Point  
$BR(\chi^0 _2 \rightarrow \tilde{l}_i l_i) \sim 0.12$ and is 
reduced due to the large branching fraction  
$BR(\chi^0 _2 \rightarrow h \chi^0 _1)$. 
\begin{table}
\begin{center}
\begin{tabular}{cccccc} \hline
$\tilde{g} \rightarrow q \tilde{q}_L $ & 30 &
$\tilde{g} \rightarrow q \tilde{q}_R $ & 30 &
$\tilde{g} \rightarrow \tilde{t}_1 t $ & 14 \\ \hline  
$\tilde{g} \rightarrow \tilde{b}_L b $ & 15 & 
$\tilde{g} \rightarrow \tilde{b}_R b $ & 10 &
$\tilde{t}_2 \rightarrow Z_0  \tilde{t}_1$ &26 \\ \hline 
$\tilde{t}_2 \rightarrow \chi^0 _4 t$ &21 &
$\tilde{t}_2 \rightarrow \chi^{\pm} _2 b$ &18  & 
$\tilde{t}_2 (\tilde{t}_1) \rightarrow \chi^{\pm} _1 b$ &15 (63)
\\ \hline 
$\tilde{t}_2 (\tilde{t}_1) \rightarrow \chi^0 _2 t $ & 8 (17) &
$\tilde{t}_2 \rightarrow \chi^0 _3 t $ & 6 &
$\tilde{t}_2 (\tilde{t}_1)\rightarrow \chi^0 _1 t $ &  6 (20) \\ \hline  
$\tilde{q}_L \rightarrow q \chi^0 _2 $ & 32  &   
$\tilde{q}_L \rightarrow q  \chi^{\pm} _1 $ & 64 & 
$\tilde{q}_L \rightarrow q  \chi^0 _1$ & 1.5 \\ \hline 
$\tilde{q}_L \rightarrow q  \chi^{\pm} _2$ & 1.5 & 
$\tilde{q}_L \rightarrow q  \chi^0_4$ & 1 & 
$\tilde{q}_R\rightarrow q \chi^0 _1$ & 99 \\ \hline 
$\chi^0 _2 \rightarrow \tilde{l}_R l $ & 36 &   
$\chi^0 _2 \rightarrow h \chi^0 _1$ & 63 & 
$\tilde{l}_R \rightarrow l \chi^0 _1 $ &100 \\ \hline 
$\tilde{l}_L \rightarrow \chi^0 _1 e $ & 90 & 
$\chi^+_1 \rightarrow W^+ \chi_1^0 $ & 98 & 
$h \rightarrow \tau \tau$ & 5 \\ \hline
\end{tabular}
\end{center}
\caption{Branching fractions (in percent) for sparticles at LHC Point 
\protect\cite{ian}. 
Here $\tilde{q}=\tilde{u}$, $\tilde{d}$, $\tilde{c}$,
$\tilde{s}$, and $\tilde{l} =\tilde{e}$, $\tilde{\mu}$, $\tilde{\tau}$.} 
\label{br}
\end{table}
This gives from
decays of first two generation
squarks 
the number of $\chi^0 _2$ pairs {\it produced}:
\begin{equation}
N \left( \chi_2^0 \chi_2^0 \right) =(0.32)^2 \times 
(\sigma_{\tilde{g} \tilde{g}} 
\times(0.3)^2 +  \frac{1}{2} 
\sigma_{\tilde{g} \tilde{q}} \times 0.3 
+\frac{1}{4} 
\sigma_{\tilde{q} \tilde{q}'} 
+\frac{1}{2}\sigma_{\tilde{q}^{\ast} \tilde{q}}) L 
\approx 3400.
\end{equation}
(The factors of $1/2$ and $1/4$ are easy to understand: $1/2$ of all 
$\tilde{q} \tilde{q}^{\ast}$ produced from s-channel gluon and 
4--point contact interaction \footnote{It is assumed that
all of the $\tilde{q} \tilde{q}^{\ast}$ production is by this channel.
This is reasonable since most of the hard collisions at LHC energies are
likely to be gluon-gluon.}, 
and $1/4$ of all 
$\tilde{q} \tilde{q}'$s produced (from t-channel gluino exchange) 
are left-handed pairs.)
This is for one year of running at low luminosity ($L =
10$ fb$^{-1}$) and for one 
detector. Hereafter estimates of event numbers will use this
integrated luminosity.
A realistic detection 
efficiency of 
$90 \%$ for single $e$, $\mu$,
and 
$90 \%$ for the single-prong decay $\tau \rightarrow \pi \nu \; 
(BR \approx 0.11)$ 
will be used. These are needed to determine the 
number of 4-lepton and 3-lepton $+ \tau$--jet events that are detected.
Later, a comment on a more realistic $\tau$-jet
detection will be made.

Next the 4--lepton signal and background are estimated.

Due to the decay chain
\begin{eqnarray}
\chi_2^0 & \rightarrow & h \chi_1^0 \nonumber \\
h & \rightarrow & \tau \tau
\end{eqnarray}
the {\em effective} $BR \left( \chi_2^0 \rightarrow \tau \tau 
\chi^0_1 \right) 
\equiv R_{\tau}$ is
\begin{eqnarray}
R_{\tau} & = & BR \left( \chi_2^0 \rightarrow \tilde{\tau}_R \tau \right)
+ BR \left( \chi_2^0 \rightarrow h \chi_1^0 \right) 
\times BR \left( h \rightarrow \tau \tau \right) \nonumber \\
 & = & 0.15.
\end{eqnarray}
Using the above BR and
$BR \left( \chi_2^0 \rightarrow \tilde{l}_R l \right) = 0.12$  
for each of $l = e$, $\mu$ and 
$BR \left( \tau \rightarrow e \nu\right) \approx BR \left( 
\tau \rightarrow \mu \nu \right)
\approx 1/2 \times 0.35$, 
\begin{eqnarray}
BR \left( \chi_2^0 \rightarrow e e \chi^0 _1,\mu \mu \chi^0 _1 \right) 
&=&
2 \times 0.12 \times \left( 1 - 2 \sin^2 \theta \cos ^2 \theta x \right)
\nonumber \\ 
& &+ R_{\tau} 
\times (0.35)^2 \times \frac{1}{2}, \nonumber \\
BR \left( \chi_2^0 \rightarrow e \mu \chi^0 _1 \right) &=&
2 \times 0.12 \times 2 \sin^2 \theta \cos ^2 \theta x + 
R_{\tau} \times (0.35)^2 \times \frac{1}{2},
\end{eqnarray}
where the first terms in each equation are from decays of
$\tilde{e}$ and $\tilde{\mu}$ and the second terms are from
$\tau$ decays.

Then, 
the total number of $4$-lepton events expected from $\chi_2^0$
pair decays (including detection efficiencies, but parameterizing 
the acceptance cut as $\varepsilon_{CUT}$ 
-- see later \footnote{Detection efficiency
refers to the probability that the lepton (or $\tau$-jet in a later case)
will be detected {\em given} that it passes the acceptance cuts. })
is
\begin{eqnarray}
N(4l) & = & N \left( \chi_2^0 \chi^0_2 \right) \times
\left(BR \left( \chi_2^0 \rightarrow e e ,\mu \mu , e \mu \right) \right) ^2
(0.9)^4 \times \varepsilon_{CUT} \nonumber \\ 
 & = & 3400 \times \left( 0.24 + R_{\tau} 
\times (0.35)^2 \right)^2 
\times (0.9)^4  \times \varepsilon_{CUT} \nonumber \\
 & \approx & 149 \times \varepsilon_{CUT}.
\label{4l}
\end{eqnarray}
To get $3e \; 1 \mu + 3 \mu \; 1 e$ events, one $\chi_2^0$ has to decay
into $e e / \mu \mu$ and the other to $e \mu$.
Thus, the number of $3e \; 1 \mu + 3 \mu \; 1 e$
events from flavor-mixing for $\sin \theta = 0.2$ and $x \sim 1$ (it is
shown later that these values are consistent with the $\mu \rightarrow 
e \gamma$
limit)
is
\begin{eqnarray}
S_{FV} & = & N \left( \chi_2^0 \chi^0_2 \right) 
\times BR
\left( \chi_2^0 \rightarrow e e \chi^0 _1, \mu \mu \chi^0 _1 \right)
\times (0.9)^4 \times \varepsilon_{CUT} \nonumber \\
& & \times 2 \times 0.24 
\times
  2 \sin^2 \theta \cos ^2 \theta x 
\nonumber \\
 & \approx & 20 \times \varepsilon_{CUT} .
\label{mixing}
\end{eqnarray}
There is an extra factor of $2$ since either $\chi_2^0$ can
decay to flavor violating dileptons. Next, the  
number of $3e \; 1 \mu + 3 \mu \; 1 e$ events from
leptonic decays of $\tau$s produced from
$\chi_2^0$ is 
\begin{eqnarray}
B_{FV} & = & N \left( \chi_2^0 \chi^0_2 \right) \times 2 \times
\left( R_{\tau} \times (0.35)^2 \times \frac{1}{2} \right) 
\nonumber \\
 & &\times
BR 
\left( \chi_2^0 \rightarrow e e \chi^0 _1,  \mu \mu \chi^0 _1 \right) 
  \times (0.9)^4 \times \varepsilon_{4l} \times \varepsilon_{CUT} \nonumber \\
 & \approx & 9 \times \varepsilon_{CUT} .
\label{tau4l}
\end{eqnarray}
Here, $\varepsilon_{4l}$ is the acceptance for $4$ leptons 
with $2$ of them coming from the decay chain $\chi_2^0
\rightarrow \tau \tau$ {\em relative} to that
for all $4$ leptons coming from $\chi_2^0 \rightarrow \tilde{e} e$
or $\mu \tilde{\mu}$. 
Since the leptons from the $\tau$ decay
are softer, it is expected that $\varepsilon_{4l} \stackrel{<}{\sim} 1$.
\footnote{Strictly speaking, the factor 
$\varepsilon_{4l}$ should be included in determining
$N(4l)$ and $S_{FV}$ as well. But since the number of events in these samples
from $\tau$ decays is very
small, it is a good approximation to 
assume $\varepsilon_{4l} \approx
1$ in those numbers.} 
To get the number in the last line above, $\varepsilon_{4l} \approx
1$ has been assumed.

Finally, the above $2$ numbers are from the leptonic decays
of $2$
$\chi_2^0$s from the decays of first two generation
squarks only. As mentioned before, 
stop/sbottom decays to $W$, $\chi_3^0$ etc.
can give a background
to the flavor violating $4$-lepton signal (see Eqn.(\ref{stopdecay})).
To reject these events, a $b$-jet veto is used. This implies that 
events with $4$ leptons coming from 
$2$ $\chi_2^0$ decays with (at least) one $\chi_2^0$ 
coming from a stop/sbottom decay will also be rejected; this is the reason
for not including the $\chi_2^0$ pairs from stop/sbottom decays in
the numbers above.

Measuring the background 
from $\chi_2^0 \rightarrow \tau \tau$ decays is discussed next.

As mentioned earlier, the idea
is to measure the number of 
$(3 e \; \tau -\hbox{jet})+  \; (2 e \; 1 \mu \;\tau -\hbox{jet})+ \cdots$
events where $\tau$-jet refers to the
hadronic decay of $\tau$. At this 
LHC Point 
the number of these events (including detection efficiencies)
is 
\begin{eqnarray}
N(3l + \tau-\hbox{jet}) & = & N \left( \chi_2^0 \chi^0_2 \right) \times 
2 \times \left( R_{\tau} \times 2 \times
0.35 \times \varepsilon_{\tau} \right) \times \varepsilon_{CUT} 
\times \nonumber \\
 & & BR 
\left( \chi_2^0 \rightarrow e e \chi^0 _1, \mu \mu \chi^0 _1, 
e \mu \chi^0 _1 \right) \times (0.9)^3 \times \varepsilon_{3l}.
\label{taujet}
\end{eqnarray}
A factor of $2$ is due to {\em either} $\tau$ decaying to a jet.
Here, $\varepsilon_{\tau}$ includes  
$BR
\left( \tau \rightarrow \hbox{hadron} \right)$ {\em and} the 
efficiency for detecting
a hadronic decay of $\tau$. The variable $\varepsilon_{3l}$ is 
the acceptance
for $(3 + 1 \tau-\hbox{jet})$ {\em relative}
to that for $4$ leptons all of which come from the decay chain $\chi_2^0
\rightarrow \tilde{e} e$, $\tilde{\mu} \mu$. It is expected
that $1 \stackrel{>}{\sim} 
\varepsilon_{3l} \stackrel{>}{\sim} \varepsilon_{4l}$
since the lepton from the $\tau$ decay is softer than the $\tau$-jet
and since the $\tau$ decay products
(both lepton and jet) are softer than the leptons from the decay chain
$\chi_2^0
\rightarrow \tilde{e} e$, $\tilde{\mu} \mu$.
From Eqns.(\ref{tau4l}) and (\ref{taujet})
and assuming $BR \left( \chi_2^0 \rightarrow e e \chi^0 _1, \mu \mu 
\chi^0 _1 \right)
\approx BR \left( \chi_2^0 \rightarrow e e \chi^0 _1, \mu \mu \chi^0 _1, 
 e \mu  \chi^0 _1 \right)$, 
the following relation is
obtained 
\begin{eqnarray}
B_{FV} & \approx & N(3l + \tau-\hbox{jet})
\times \frac{0.9 \times \frac{0.35}{2} \times
\varepsilon_{4l}}{2 \times \varepsilon_{3l} \times \varepsilon_{\tau}}.
\label{4l3ljetrel}
\end{eqnarray}
Note that $R_{\tau}$ {\em cancels} in the ratio. 
Thus,
using
the $\left( 3l + \tau-\hbox{jet} \right)$
detection together with an understanding 
of the $\tau$ detection efficiency ($\varepsilon_{\tau}$), 
as well 
as the acceptance for 4 leptons (with $2$ of them from $\tau$ decays)
versus (3 leptons $+ \tau$-jet) ($\varepsilon_{4l} / \varepsilon_{3l}$), 
the number of $(3e\; 1\mu+ 3 \mu \; 1e)$ events 
from $\tau$ decay 
(Eqn.(\ref{tau4l}))
contained in the full 4-lepton sample can be obtained
from the above relation.
This is important, as it means that
the $\chi^0 _2 \rightarrow \tau \tau$ background
to the flavor-violating signal
can be determined without knowing the relative branching fraction
of $\chi^0 _2$ to $h$,
$\tilde{l}l$, or $\tilde{\tau} \tau$.

Assuming that 
the detection efficiency for the decay
$\tau \rightarrow \pi \nu$ (which has a BR of 0.11)
is $0.9$ so that
$\varepsilon_{\tau} \approx 0.9 \times 0.11$,
and assuming $\varepsilon_{3l} \approx 1$ gives 
\begin{equation}
N(3l + \tau-\hbox{jet}) \approx 11 \times \varepsilon_{CUT}.
\end{equation}


Independent of this, it is worth remarking that with enough 
stastistics it might be possible to measure 
$BR(\chi^0 _2 \rightarrow  h \chi^0 _1)$, 
$BR(\chi^0 _2 \rightarrow  \tilde{e} e, \tilde{\mu} \mu)$ and 
$BR(\chi^0 _2 \rightarrow \tilde{\tau} \tau)$ {\em assuming} that
these are the dominant decay modes of $\chi_2^0$. The decay chain 
$\chi^0 _2 \rightarrow h \chi^0_1 \rightarrow b \bar{b} \chi^0_1$
(where $\chi_2^0$ is from cascade decays of
squarks as usual) gives $b \bar{b}$ events with high
$p_T$ jets and $\not \! p_T$. Comparing these to  
the number of dilepton events from $\chi^0 _2$ decays gives
\begin{equation}
\frac{N(b \bar{b})}{N(2l)} \propto 
\frac{BR(\chi^0 _2 \rightarrow h \chi^0 _1)}{BR(\chi^0 _2 \rightarrow 
\tilde{e} e \chi^0 _1, \tilde{\mu} \mu \chi^0 _1)}.
\end{equation}
Similarly, the number of  
$3l+1\tau$--jet events compared to 4--lepton events is 
\begin{equation}
\frac{N(3l+\tau- \hbox{jet})}{N(4l)} \propto 
\frac{R_{\tau}}
{BR(\chi^0 _2 \rightarrow 
\tilde{e} e \chi^0 _1, \tilde{\mu} \mu \chi^0 _1)}.
\end{equation}
All the events in the above two equations have in addition
high $p_T$ jets and $\not \! p_T$ to make sure that these are from 
cascade decays of squarks.
From these two measurements and the assumption that $\sum BRs=1$ the 
above--mentioned branching ratios can be obtained. This could provide 
complementary information to the flavor violating signal discussed here.  
    
Returning to the main subject of this section, 
an observation of an
excess 
of the `flavor violating'
4--leptons events 
over those from $\tau$ decay (Eqn.(\ref{tau4l}))
would be a strong evidence for lepton
flavor violation.
But, before concluding that SUSY lepton
flavor violation has been detected, the background to the
flavor violating $4$ lepton events from stop/sbottom
production (see Eqn.({\ref{stop}) below) must be removed, 
and also
the $\tau$-jet detection efficiency $\varepsilon_{\tau}$ must be known.
These two issues are discussed next.

The $\tau$ hadronic decays from $Z \rightarrow \tau \tau$ at the 
LHC were
simulated for the ATLAS detector
in \cite{atlas} \footnote{There is also a study of
detecting $\tau$-jets from heavy SUSY Higgs decay for the CMS detector
\cite{cms}.}. This study 
shows that
a detection efficiency $\epsilon_{\tau}$ for
a hadronic $\tau$ decay (including the multi-prong decays, {\it i.e,}
a total $\tau$ decay BR of 0.65) of $\approx 40 \times 0.65 \%$
with a rejection factor of 15 for non-$\tau$ jets
can be achieved. This is possible since
$\tau$-jets have lower particle multiplicity, narrower profile and 
smaller invariant mass than the QCD jets \cite{atlas}. 
A similar detection efficiency (or even better detection
efficiency and rejection of non-$\tau$ jets
if the strategy is optimized for this case)
for $\tau$-jets from sparticle decays could be expected.

The important point about this though is that it
suffices to know a {\em lower} limit on the $\tau$-jet detection efficiency
to get an {\em upper} limit on the number of $\left( 3 e \; 1 \mu \right) + 
\left(3 \mu \; 1 e
\right)$ events from tau decays using
the $\left( 3 e \; \tau -\hbox{jet} \right)$ 
events (see Eqn.(\ref{4l3ljetrel})). Similarly, since 
$\varepsilon_{4l} \stackrel{<}{\sim} \varepsilon_{3l}$,
an {\em upper} limit on $B_{FV}$ can be obtained even
though these $\varepsilon$'s may not be known precisely.
Also, if the $\tau$-jet detection (and QCD jet rejection) is good,
there will be large number of events with 
$2$ leptons and $2$ $\tau$-jets from $2$ $\chi_2^0$ decays.
These can be used in addition to the $3$ lepton $1 \tau$-jet 
events to estimate the background to flavor violating $4$ lepton events
from $\tilde{\tau}/\tau$ decays.

To reduce the stop and sbottom backgrounds a $b$-jet veto 
can be used. Before using this veto, the number of 
expected $3e \; 1\mu+3 \mu \; 1e$ events from decays of 
$\tilde{t}$ or $\tilde{b}$ to $W$, $\chi^0 _3$, $\chi^{\pm}_1$ etc.
(in the absence of any flavor mixing) 
can be estimated using the production cross-sections and 
branching fractions from Tables \ref{cross-section} and \ref{br}. 
The result is, including lepton detection efficiencies:
\begin{equation}
 N(\tilde{t} \; \hbox{or} \; \tilde{b})\approx 50 \times \varepsilon_{CUT}. 
\label{stop}
\end{equation} 
Each of these events has at least
$2$ $b$ quarks. So with a $b$-detection 
efficiency of
$60 \%$ (and rejection factor of $200$
against non-$b$ jets at low luminosity
\cite{tp}), the number of $3 e \; 1 \mu + \cdots$ events 
from stop/sbottom decays after the $b$-jet veto
goes down to $8$. This can be further reduced 
by using a $b$-tagging efficiency of $90 \%$ 
with a mistag rate of $25 \%$ ({\it i.e.}, rejection factor
of $4$ against non-$b$ jets) at low luminosity \cite{ian,tp}; this 
will reduce the signal by a bit. This  
strategy can be optimized 
depending on the luminosity \cite{tp}.

Lastly, to get actual number of
events, the cuts used to select these events must also 
be taken into account. The effect of these 
cuts on the signal and background rates is buried in the  
fudge factor $\varepsilon_{CUT}$. 
For example, $p_T \stackrel{>}{\sim} 10$ GeV and $\mid \eta \mid
\stackrel{<}{\sim} 2.5$ is required 
to be able to detect $e$ or $\mu$.
Also, to reduce any 
remaining small SM background,
{\it i.e.,} to make sure that these {\em are} SUSY events, various 
cuts on
$\not \! p_T$, $p_T$ of jets, a variable $M_{eff}$ \cite{snowmass,ian}
related
to $\not \! \! p_T$, $p_T$ of jets, can be imposed.  
Analysis of
the events simulated in
\cite{ian} showed that there were $\sim 40$
events with $4$ leptons with no $b$-jets
that pass all the cuts mentioned
above compared to the estimate of $\sim 149$ from  
cross sections and BRs, Eqn.(\ref{4l}): there is an 
acceptance factor of $\varepsilon_{CUT} \sim 1/4$ 
from the various kinematic cuts.
We have also checked that almost all of these 
(simulated) events have $2$ $\chi_2^0$s as expected.
\footnote{The information about whether an event in the simulation
has
$\chi_2^0$s,
$\tilde{t}$s, $\chi_3^0$s etc. is from the event generator.}
There are very few events in this sample (from the simulation)
with heavier neutralinos/chargino in agreement with the 
expectation from the very small
BRs of the first two generation squarks to these sparticles at 
this point in the SUSY parameter space \cite{ian} (see Table \ref{br}).
The number of events (from the simulation)
with at least $1$ $b$ quark and $4$ leptons
is also in rough agreement (up to the acceptance factor) with the 
number of $4$ lepton events with
at least $1$ stop/sbottom
expected from the cross section and branching fraction 
estimates. \footnote{We have also checked that
these simulated events {\em do}
have at least $1$ stop/sbottom.
There are very few events in this sample with {\em no}
stops/sbottoms but with $b$-jets from initial state
gluon radiation.}

Including an acceptance factor of $\varepsilon_{CUT} \sim 1/4$ for both 
background and signal, a 
$b-$jet detection efficiency of 60$\%$ (which was not included in 
Eqn.(\ref{stop}))
and 
detection efficiency of 90$\%$ for
the decay $\tau \rightarrow \pi \nu$, and a 66$\%$ 4-lepton detection 
efficiency (the $\tau$ and lepton detection efficiencies were
included in the previous estimates of $S_{FV}$ etc.), 
a summary of the expected number of 
events at {\it low luminosity} is :
\begin{eqnarray}
N (4l) & \approx & 37 \times \frac{L}{\hbox{10 fb}^{-1}} \nonumber \\
S_{FV} & \approx & 5 \times \frac{\sin^2 2 \theta}{0.15} x \times 
\frac{L}{\hbox{10 fb}^{-1}}  \nonumber \\
B_{FV} & \approx & 2 \times \frac{L}{\hbox{10 fb}^{-1}}  \nonumber \\
N(3l+\tau-\hbox{jet}) & \approx 
& 3 \times \frac{L}{\hbox{10 fb}^{-1}}  \nonumber \\
N(\tilde{t} \hbox{ or }\tilde{b}) & \approx 
& 2 \times \frac{L}{\hbox{10 fb}^{-1}} .
\end{eqnarray}
While these numbers may be a little small for  
one detector and 
one year of running at low luminosity $(L=10$fb$^{-1}$), there is cause for 
optimism. 
More {\em integrated} luminosity $L$ from $>1$ year
of running and/or $2$ detectors can significantly increase the statistics. 
Further, a 
 larger $BR(\chi^0_2 \rightarrow 
\tilde{l} l)$ would give more statistics. 
This could occur at a point in the SUSY parameter space with a heavier 
Higgs boson, and thus a lower   
$BR(\chi^0_2 \rightarrow h \chi^0_1)$. 

To illustrate the discovery or exlcusion significance of these 
results, an integrated luminosity of $L=100$ fb$^{-1}$ is 
considered.  
This could occur for 
5 years of running at low luminosity for two detectors 
\footnote{One year of running at high luminosity is also possible. In this 
case however, the $b-$jet mistag rate increases to 1 in 6 for a 
$b-$tagging efficiency of 80$\%$ \cite{tp}. Since most of the signal events 
occur from $\tilde{q} \tilde{g}$ production and so contain at least 
three 
hard jets, approximately 40$\%$ of the signal could be rejected. 
In this case the discovery (and exclusion) 
limits on $\sin \theta_R$ increase by about 
$25\%$. In addition, the tau-jet detection efficiency at high 
luminosity is not known since a low luminosity was used in 
the ATLAS study.}.
For this integrated luminosity there are 22 4--lepton flavor 
violating events from the $\tilde{\tau}/\tau$ background, and 
30 3--lepton $\tau-$jet events. There will also be 
125 4--lepton flavor
violating events {\it before the $b-$jet veto} from the 
$\tilde{t}/ \tilde{b}$ background. Next, 
the $b-$tagging efficiency is optimized so that 
the $\tilde{t}/ \tilde{b}$ background is (less than or)
equal 
to the $1\sigma$ statistical error in $\tilde{\tau}/\tau$ background
while at the same time the reduction of the signal due to mistagging
is small. 
This is achieved with a $b-$tagging efficiency of 
$80 \%$, rather than the $60 \%$ of before. At this higher tagging 
efficiency there is a mistag rate of 1 in 50, so there is very little
reduction of the signal. 
With an $80 \%$ $b-$tagging efficiency, 5 
$\tilde{t}/ \tilde{b}$ background events remain since each event has 
at least 2 $b-$jets. Then the background is dominated by the 
$\tilde{\tau}/\tau$ decays.
A $5\sigma$ $(2\sigma)$ discovery 
(exclusion) requires that 
 $S/\sqrt{B} >5 $ $(S/\sqrt{B} > 2)$, and this requires 
$>$23 ($>9$) signal events. So a $5\sigma$ discovery is obtained for 
\begin{equation}
 \sqrt{x} \sin 2 \theta_R  > 0.26 \;  (5 \sigma \; \hbox{discovery}) 
\; \hbox{ or } 
   \sin \theta_R > 0.13 \; \hbox{ for } x\sim 1. 
\end{equation}
If no signal is observed then the $2 \sigma$ exclusion limit is  
\begin{equation}
\sqrt{x} \sin 2 \theta_R  > 0.16 \; (2 \sigma \; \hbox{exclusion}) 
\; \hbox{ or }
   \sin \theta_R > 0.08 \; \hbox{ for } x\sim 1.
\end{equation}

To end this Section, these values of $\sin 2 \theta_R$ and $\Delta m/m$ that 
may be probed by the LHC are compared to the constraints on these 
parameters obtained from 
$\mu \rightarrow 
e \gamma$. 
The LHC signal is proportional to 
$\sin ^2 2 \theta _R \; x$, with $x \sim 1$ if $\Delta m \stackrel{>}{\sim}
\Gamma$ and  
$x \ll 1$ if $\Delta m \ll \Gamma$. The decay $\mu \rightarrow
e \gamma$ 
places an upper limit on 
$\sin 2 \theta _R \Delta m /m$ (Eqn.(\ref{muegamma}))
so that there is competition between the two probes of flavor
violation.
Thus, in order for the signal at the LHC to be significant in 
the region of the
$\left( \sin 2 \theta _R, \; \Delta m /m \right)$ plane {\em beyond}
the reach of the 
$\mu \rightarrow e \gamma$ limit, 
there should be a range of $\Delta m /m$
where $\Delta m \stackrel{>}{\sim} \Gamma$ 
so that $x \sim 1$ {\em and} $\Delta m/m$ is small enough
(for a given value of $\sin 2 \theta_R$)
so that $\mu \rightarrow
e \gamma$ is suppressed.  
It will be seen that for  
$\Delta m/m \sim \Gamma /m$ (so that
$x \sim 1$), 
at this LHC Point,  
$\sin 2 \theta_R$ is unconstrained by the $\mu \rightarrow e \gamma$ 
limit, affording the LHC the opportunity to either detect a signal 
or extend the limit.   

At this Point $\chi^0 _1 \approx \tilde{B}^0$. A computation 
of the one-loop 
$\tilde{B}^0$ contribution gives  
\begin{equation}
\sin 2 \theta _R \frac{\Delta m^2_R}{\tilde{m}_R ^2}
\left(\frac{\hbox{100 GeV } \tilde{m}_R}{M^2 _{\chi^0 _1}} \right)^2
20 F(\alpha_L, \alpha_R,t)  <
0.013 \times
\sqrt{\frac{BR(\mu \rightarrow e \gamma)}{4.9 \times 10^{-11}}}.
\label{ueg}
\end{equation}
Here $\alpha_H=\tilde{m}^2 _H /M^2 _{\chi^0 _1}$ $(H=L,R)$, 
$t=(A+\mu \tan \beta )/ \tilde{m}_R$,
\begin{equation}
F(\alpha_L, \alpha_R,t)=H(\alpha_R)+t \alpha^{1/2} _R
\frac{\partial K}{\partial \alpha_R}(\alpha_L,\alpha_R),
\end{equation}
with
\begin{equation}
K(x,y)=\frac{g(x)-g(y)}{x-y}, \; g(x)=\frac{1+2 x \log x- x^2}{2 (x-1)^3},
\end{equation}
and
\begin{equation}
H(x)=\frac{-x^3+9 x^2+9x-17-(6+18 x) \log x}{6 (x-1)^5}.
\end{equation}
Two useful facts are $H(1)=\frac{\partial}{\partial y} K(1,1)=-1/20$; 
hence the factor of 20 on the 
left side of Eqn.(\ref{ueg}).
At this LHC Point, $m_{\tilde{l}_R} \sim $160 GeV, 
$m_{\chi^0_1} \sim$ 120 GeV,
and $m_{\tilde{l}_L} \sim$ 240 GeV. 
Inputing these masses into the above formula simplifies it to: 
\begin{equation}
\frac{\sin 2 \theta_R}{0.39} 
\times \frac{\Delta m_R}{\tilde{m}_R} \times(1 +0.48 t) < 0.03 \times 
\sqrt{\frac{BR(\mu \rightarrow e \gamma)}{4.9 \times 10^{-11}}} .
\label{nmue}
\end{equation}
At this Point $t \approx 5-10$. However, a 
larger variation in $t$ is allowed without affecting the 
flavor violating signal, since both  
$A$ and sgn$(\mu)$ do not qualitatively affect 
the 4--lepton event rate \footnote{It is important to maintain 
the relation $m_{\chi^0 _2}> m_{\tilde{e}_R}$ though.}.   
In any case, the values $\sin 2 \theta \approx 0.39$ 
and 
$\Delta m_R \sim \Gamma$ 
(so that $x \sim 1$) with a typical
value of $\Gamma \sim \alpha_{em} m \sim 10^{-2} \times \tilde{m}_R$ 
are consistent with 
$\mu \rightarrow e \gamma$ -- recall that $\sin
2 \theta \approx 0.39$ and $x \sim 1$ was assumed to 
obtain the estimate of $S_{FV}$ in Eqn.(\ref{mixing}). 
In fact, at this LHC Point 
$\Gamma \sim 125$ MeV \cite{ian}, so that $\Gamma / m \sim 8 \times 
10^{-4}$ which is smaller than $\alpha_{em}$. 
So for $\Delta m/m \stackrel{>}{\sim}
2 \; \Gamma/ m \approx 1.6 \times 10^{-3}$, 
it follows that $x \sim 1$. From Eqn.(\ref{nmue}) and for maximal mixing 
($\sin 2 \theta _R=1$), 
$\Delta m/ m < 0.39 \times 0.03 /(1+0.48 t) \approx  4 \times 10^{-3}$ (for
$t \approx 5$). Thus 
for $1.6 \times 10^{-3} \stackrel{<}{\sim}
\Delta m/ m \stackrel{<}{\sim} 4 \times 10^{-3}$ 
and $\sin 2\theta_R =1$, $\mu \rightarrow e \gamma$ is satisfied 
and $x \sim 1$.
So at this 
Point even for maximal mixing 
there is a large range of $\Delta m/m$ for which 
$x \sim 1$ and $\mu \rightarrow e \gamma$ is safe. Of course, smaller 
mixing could be probed by the LHC, in which case the upper 
bound on $\Delta m/m$ allowed by $\mu \rightarrow
e \gamma$ is larger. In this case 
for a given $\sin \theta _R$ there is a larger range of
$\Delta m /m$ 
for which $x \sim 1$ (so that there is no
suppression of the LHC signal)
{\em and} $\mu \rightarrow e \gamma$ is
safe. 

\section{Conclusions}
We believe that it is possible
to detect SUSY lepton flavor violation 
at the LHC using events with $4$
leptons from the cascade decays of squarks provided the following
conditions are satisfied:
 
0. Either $R-$parity is conserved or $\chi^0 _1$ $(LSP)$ decays 
outside the detector,  

1. $\chi_2^0$ pair production in cascade decays of squarks is large and
$\chi_2^0$ has a large decay branching fraction 
to $\tilde{l}^{\ast} l$ (to get 
enough statistics),

2. Hadronic decays of $\tau$s can be detected with a known efficiency
so that the background from the $\chi_2^0 \rightarrow
\tau \tau$ decay can be estimated,

3. The $b$-jet detection
efficiency is good so that the background from 
events with stop/sbottom 
can be rejected,

4.  
The stop/sbottom production rate, either 
direct or in gluino decays, is not {\em too much} larger than the 
production of
first two generation squarks,   
 
5. The first two generation squarks  
decay largely to $\chi_2^0$, $\chi_1^0$ and 
$\chi_1^{\pm}$, so that the background to flavor violating $4$
lepton events from decays of heavier neutralinos/chargino 
to $W$s, lighter chargino, {\em heavier}
sleptons etc. is small.
This condition can be realised in the so-called gaugino-like region, 

6. The mass splitting is $\Delta m \sim \Gamma$ or {\em larger}, 
so there is no
suppression of the signal due to the quantum interference effect.  

The arguments presented here are clearly semi-quantitative, and 
further study requiring a detailed
simulation of these processes is required.
This is beyond the scope of this work. 

\section{Acknowledgements}
We thank Nima Arkani-Hamed and Mahiko Suzuki 
for discussions, and Ian Hinchliffe in particular for discussions 
and allowing us to use 
the simulations 
of Point 5 of
the LHC study. This work was
supported in part by the U.S. Department of Energy under Contract
DE-AC03-76SF00098.
KA thanks 
Frank Paige for
discussions. MG thanks the Natural Sciences and Engineering Research 
Council of Canada for their support.


\begin{thebibliography}{99}

\bibitem{krasnikov0} N.V. Krasnikov, Phys.Lett. {\bf B388}, 783 (1996). 
\bibitem{nima1}N. Arkani-Hamed {\it et al.}, Phys. Rev.
Lett. {\bf 77}, 1937
(1996). 
\bibitem{nima2} N. Arkani-Hamed {\it et al.}, Nucl. Phys. {\bf B505}, 3 (1997).
\bibitem{krasnikov2}N.V. Krasnikov, JETP Lett. {\bf 65}, 148 (1997).
\bibitem{krasnikov1} S. I. Bityukov and N.V. Krasnikov, hep-ph/9712358.
\bibitem{flavor} See, for example,
N. Arkani-Hamed, H.--C. Cheng and L.J. Hall, Nucl. Phys. {\bf B472}, 
95 (1996), and references therein.
\bibitem{baer}H. Baer {\it et al.}, Phys. Rev. {\bf D49}, 3283 (1994).
\bibitem{snowmass}A. Bartl {\it et al.},
Proceedings of the 1996
DPF Summer Study (Snowmass, CO, 25 June - 12 July, 1996), LBNL-39413 (1996).
\bibitem{ian}I. Hinchliffe {\em et al.}, Phys. Rev. {\bf D55}, 5520 (1997)
and private communication with I. Hinchliffe.
\bibitem{paige}F. Paige, communication at the conference.
\bibitem{atlas}Y. Coadou {\it et al.}, 
ATLAS Internal Note ATL-PHYS-98-126, LBNL-41757.
\bibitem{cms}See p.191-192 of chap.12 (Physics
Performance) of the CMS Technical Proposal at 
http://cmsinfo.cern.ch/TP/TP.html.
\bibitem{tp}See Fig.3.42, p.98 of
ATLAS Collaboration, Technical Proposal, CERN/LHCC/94-43,
LHCC/P2 (1994).

\end{thebibliography}
\end{document}